\documentstyle[12pt,epsfig]{article}

\def\be{\begin{equation}}
\def\ee{\end{equation}}
\def\ba{\begin{eqnarray}}
\def\ea{\end{eqnarray}}
\def\ga{\mathrel{\raise.3ex\hbox{$>$\kern-.75em\lower1ex\hbox{$\sim$}}}}
\def\la{\mathrel{\raise.3ex\hbox{$<$\kern-.75em\lower1ex\hbox{$\sim$}}}}

\begin{document}

\baselineskip=16pt
\begin{titlepage}
\rightline{UMN--TH--2356/05} \rightline{astro-ph/0504650}
\rightline{April 2005}
\begin{center}

\vspace{0.5cm}

\large {\bf Constraints on the dark energy equation of state from
the separation of CMB peaks and the evolution of $\alpha$.}
\vspace*{5mm} \normalsize

{\bf Seokcheon Lee}

\smallskip
\medskip

{\it School of Physics and Astronomy,\\ University of Minnesota,
 Minneapolis, MN 55455, USA}

\smallskip
\end{center}
\vskip0.6in

\centerline{\large\bf Abstract} We introduce a simple
parametrization of the dark energy equation of state, $\omega$,
which is motivated from theory and recent experimental data. The
theory is related to the tracker solution to alleviate the fine
tuning problem. Recent experimental data indicates that the
present value of $\omega$ is close to $-1$. We analyze the
evolution of $\omega$ from the separation of CMB peaks and the
time variation of the fine structure constant, $\alpha$. We find
that $ -1.00 \leq \omega^{(0)} \leq -0.971_{-0.027}^{+0.017}$ and
$ 1.76_{-0.42}^{+0.29} \times 10^{-4} \leq (d \omega / dz)_{z=0}
\leq 0.041_{-0.037}^{+0.016}$ at $95$\,\% confidence level (CL).

\vspace*{2mm}

\end{titlepage}

\section{Introduction}
\setcounter{equation}{0}

 Recent cosmological observations show that a
component with negative pressure (dark energy) should be added to
the matter component to give the critical density today
\cite{SCP}. The cosmological constant and/or a quintessence field
are the most commonly accepted candidates for dark energy. The
latter is a dynamical scalar field leading to a time dependent
equation of state, $\omega$.

 There are certain hopes that quintessencelike  models may help
alleviate the severe fine-tuning problem associated with the
cosmological constant and that the nature of dark energy can be
understood by measuring the behavior of $\omega$ with respect to
time \cite{Ratra,FJ}. But it may not be practical to test every
single quintessence model by using experimental data. So a model
independent approach for quintessence could be an effective way to
study the properties of dark energy.

 To investigate the cosmological evolution of $\omega$, we may use
the relation between the luminosity distance and redshift ($z$),
which can be obtained from type Ia supernovae (SNe Ia) as standard
candles \cite{Linder}. In this case, the distribution of the
supernovae (SNe) along $z$ provides a clue for the determination
of $\omega$. The dependency of the luminosity distance on $\omega$
increases in proportion to $z$. The separation of cosmic microwave
background (CMB) peaks can also be used to investigate $\omega$
because the location of peaks depends on the amount of dark energy
today and at last scattering as well as $\omega$ \cite{QCMB}. In
addition, the cosmological evolution of the fine structure
constant, $\alpha$ may be used to determine the $z$ dependence of
$\omega$ \cite{wz}. The advantage of using $\alpha$ is that there
will not be any degeneracies between parameters due to the
multiple integrals as in the determination of luminosity distance
\cite{Maor}.

 A number of parametrizations of $\omega$ have been used to
probe the nature of dark energy. The simplest is the averaged
value $\bar{\omega}$, defined as $\bar{\omega} = \int \omega(z)
\Omega_{\phi}(z) dz / \int \Omega_{\phi}(z) dz$ where
$\Omega_{\phi}$ is dark energy density contrast. This
parametrization is weighted by $\Omega_{\phi}$ indicating that
$\omega$ is more significant if $\rho_{\phi}$ is a larger part of
the critical density. If $\omega$ does not change during the
recent history of the universe, when $\Omega_{\phi}$ is comparable
to matter density ($\Omega_{m}$), then the average
$\bar{\omega}^{(0)}$ will be equal to present value,
$\omega^{(0)}$. The next intuitive method is a first order Taylor
expansion of $\omega$ in $z$, $\omega(z) = \omega|_{z=0} + d
\omega / dz|_{z=0} z$. This diverges at large value of $z$, which
makes it only useful for low $z$ analyses, such as SNe \cite{HM}.
Another simple parametrization is $\omega = \omega_{0} +
\omega_{1}z/(1+z)^{p}$ where $p = 1, 2$. When $p$ is equal to one,
$\omega_{0} + \omega_{1}$ is constrained as a negative value to
satisfy SNe data. In addition, $\Omega_{\phi}$ at decoupling
should be negligible based on WMAP \cite{Spergel}. However with
this constraint we can not get a small value of $\Omega_{\phi}$ at
decoupling. The $p = 2$ case can solve this problem \cite{zp}. A
simple extension of the previous cases is given by $\omega(a) =
\omega_{0}\omega_{1} (a^{q} + a_{s}^{q})/(\omega_{1} a^{q} +
\omega_{0} a_{s}^q)$, which includes huge flexibility and the
intuitive role of each parameter \cite{HM}. In this case, both
$\omega_{0}$ and $\omega_{1}$ should have the same (negative)
signs. Otherwise, $\omega$ will diverge at $a^{q} =
-(\omega_{0}/\omega_{1}) a_{c}^q$. There have also been more
complicated parametrizations \cite{parametrization}.

 We introduce a new parametrization of $\omega$, which looks
similar to that in \cite{HM}. However it is necessary that the
sign of $\omega_{0}$ is opposite to that of $\omega_{1}$ to have a
smoothly changing $\omega$. We will not consider the case of
$\omega < -1$, which can be obtained in scalar-tensor gravity
models \cite{st}, phantom models \cite{phantom}, and brane models
\cite{brane}.

 This paper is organized as follows. In the next section we introduce
a parametrization of $\omega$ and specify some of its parameters.
By changing these parameters we can mimic several quintessence
models. In section 3, we use CMB peaks to restrict the parameters.
We check the time variation of $\alpha$ based on this
parametrization in section 4. Our conclusion is in the last
section.

\section{Parametrization of $\omega$ }
\setcounter{equation}{0}

 For cosmological evolution equations, it is convenient to
introduce a variable $x$ as the logarithm of the scale factor $a$,
\be x = \ln a = - \ln(1+z) \label{x} \ee where we choose the
present scale factor $a^{(0)} = 1$. We propose a simple
parametrization of the dark energy equation of state, \be
\omega(a) = \omega_{r} \frac{\omega_{0} a^q + a_{c}^q}{a^q +
a_{c}^q} \hspace{0.2in} \Longrightarrow \hspace{0.2in} \omega(x) =
\omega_{r} \frac{\omega_{0} \exp (qx) + \exp (qx_{c})}{\exp (qx) +
\exp (qx_{c})} \label{omega} \ee where $\omega_{r}$, $\omega_{0}$,
$a_{c}$ (equally $x_{c}$), and $q$ are constants. Instead of
leaving $\omega_{r}$ and $\omega_{0}$ as arbitrary constants, we
adopt the tracking condition in the early universe $i.e.$ $\omega$
changes as that of radiation ($\omega = \omega_{r} = 1/3$). Also
we use the experimental evidence which indicates $\omega$
approaches negative one at present. These constraints fix two
constants in Eq. (\ref{omega}) : \be \omega_{r} = \frac{1}{3},
\hspace{0.2in} \omega_{0} = -3 \label{omegas} \ee

\begin{center}
\begin{figure}
\vspace{1cm} \centerline{ \psfig{file=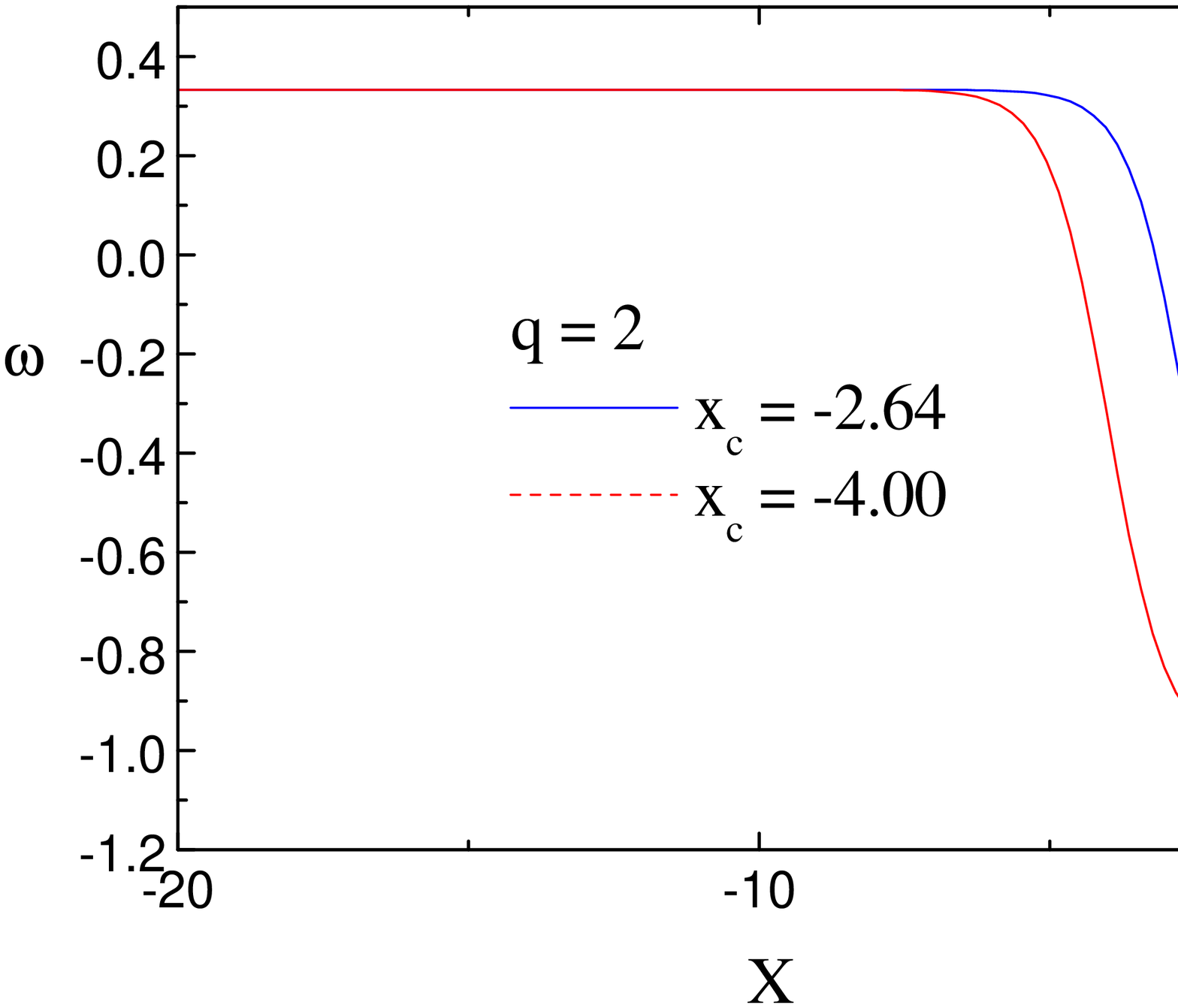,
width=8cm}\psfig{file=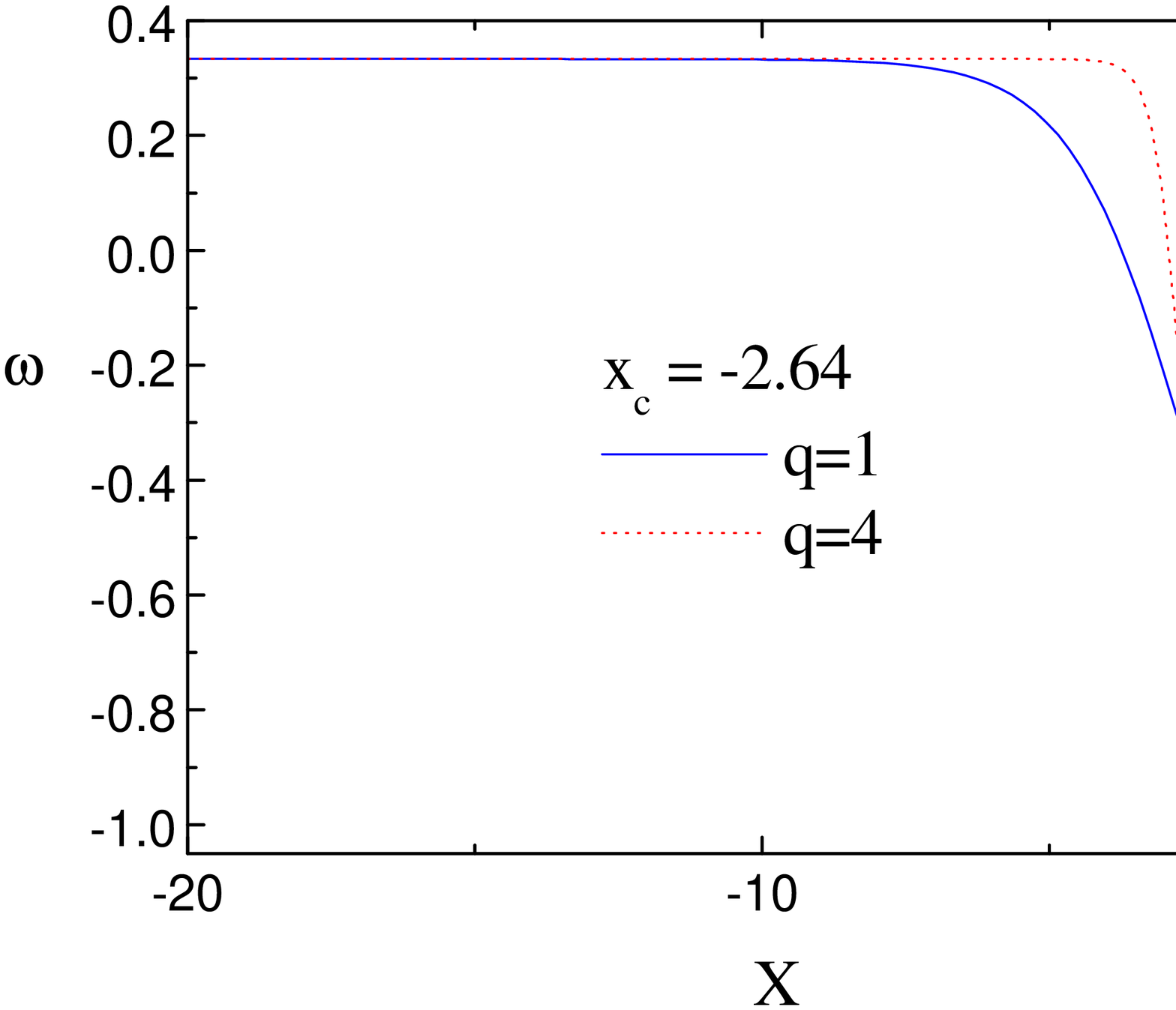, width=8cm} } \vspace{-2cm}
\caption{ The evolution of $\omega$. (a) The dependence on the
changeover scale, $x_{c}$ of the evolution of $\omega$. A smaller
value of $x_{c}$ means an earlier change of $\omega$ shown as the
dashed line. (b) The dependence on $q$ of the evolution of
$\omega$. The solid line ($q=1$) shows a slower change than the
dashed one ($q=4$). That means the duration of the changeover
becomes longer as the value of $q$ becomes smaller.}
\label{fig:fig1}
\end{figure}
\end{center}

\begin{center}
\begin{figure}
\vspace{1cm} \centerline{\psfig{file=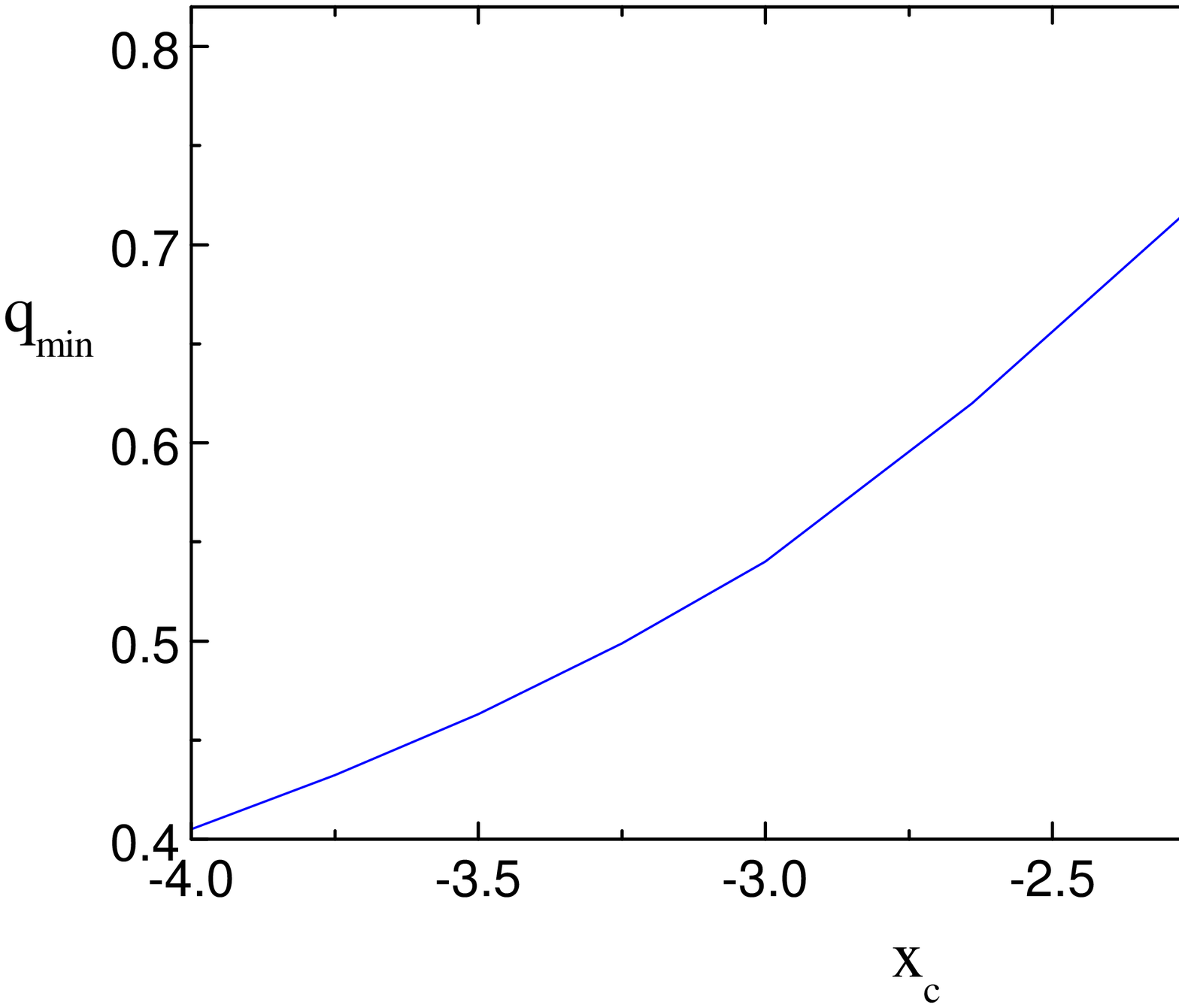, width=8cm}
\psfig{file=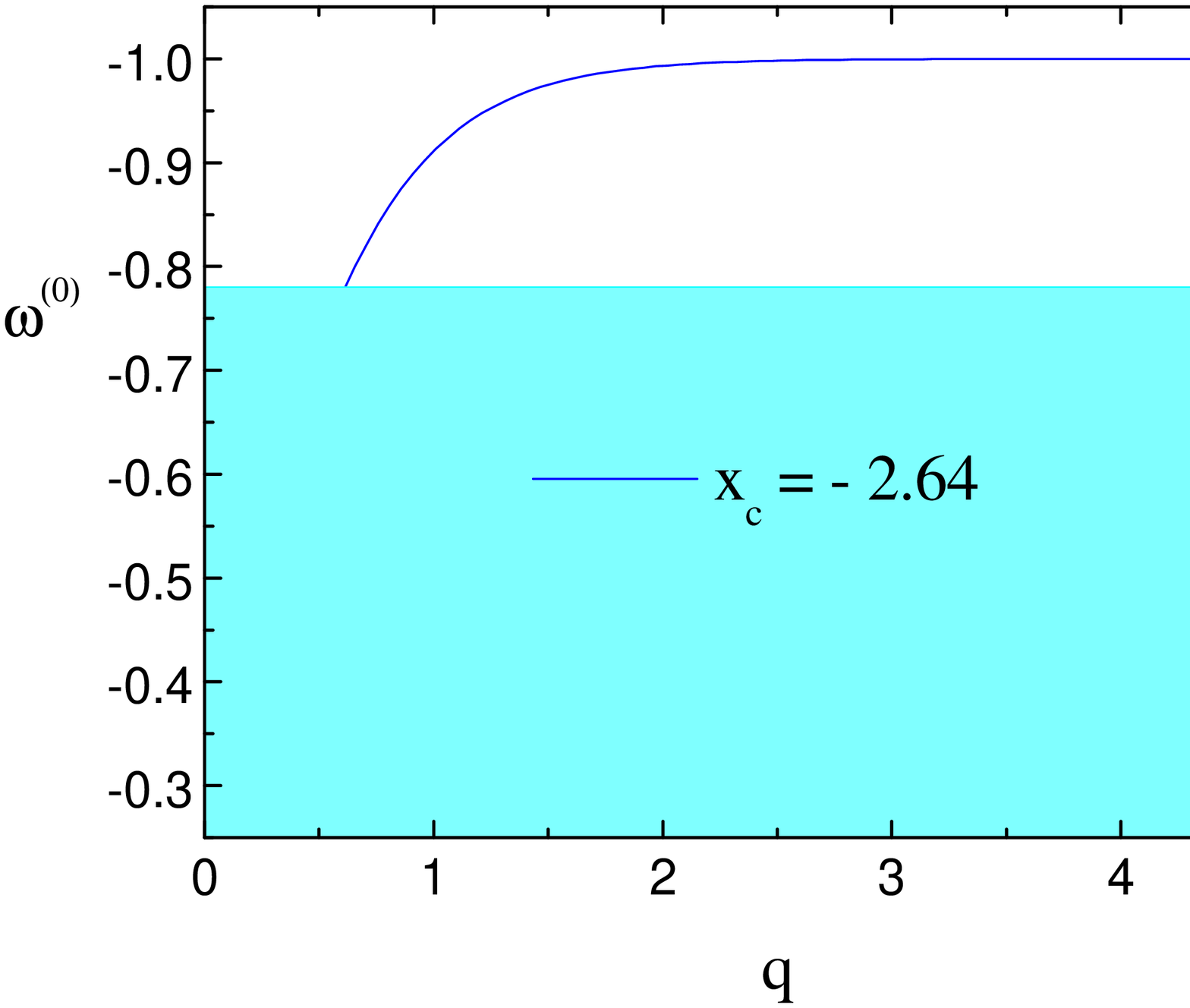, width=8cm} } \vspace{-2cm} \caption{ The
present value of the equation of state, $\omega^{(0)}$. (a)
$q_{\rm{min}}$ means the minimum value of $q$ for each different
value of $x_{c}$ to satisfy WMAP data ($\omega^{(0)} < -0.78$)
\cite{Spergel}.  (b) The shaded region is forbidden from WMAP.
From this we can make the restriction $q \geq 0.62$ when $x_{c} =
-2.64$. The reason for the choice of this specific value of
$x_{c}$ will be shown later.} \label{fig:figw0}
\end{figure}
\end{center}

There remain two arbitrary constants $a_{c}$ and $q$, which
describe the scale factor at changeover and the duration of it
respectively. When $a$ is close to $a_{c}$, $\omega$ starts to
decrease and $\omega$ reaches to $-1$ as $a$ approaches $a^{0}
=1$. We call a transition of $\omega$ from the tracking region to
the $\omega \sim -1$ region as changeover. We can also narrow the
range of the $a_{c}$ value, when we consider the fact that initial
value of $\rho_{\phi}$ is expected to be $120$ orders of magnitude
larger than the present value. We will see this in the next
section. As we can see in Eq. (\ref{omega}), for larger value of
$q$, the duration of changeover of $\omega$ decreases. During
early times $\omega \simeq \omega_{r}$ and it stays with same
value before $a$ approaches $a_{c}$. $\omega$ will approach
$\omega_{r} \times \omega_{0}$ when $a \gg a_{c}$. This is the
reason we choose the values in (\ref{omegas}). If $a_{c}$ is too
close to $a^{(0)} = 1$, then the present value of $\omega$ will
not be close to $-1$ for small $q$. Thus there is a limit to
$x_{c}$ for each value of $q$. The cosmological evolution of
$\omega$ due to this parametrization for different values of $q$
and $x_{c}$ are shown in Fig. \ref{fig:fig1} and Fig.
\ref{fig:figw0}. The effect of $x_{c}$ on the evolution of
$\omega$ is shown in Fig. \ref{fig:fig1} (a). A smaller value of
$x_{c}$ gives an earlier change of $\omega$. As $x_{c}$ decreases,
the period of $\omega = -1$ increases. A smaller value of $q$
shows a slower change of $\omega$ as in Fig. \ref{fig:fig1}(b).
For a specific value of $x_{c} = -2.64$, we require $q \geq 0.62$
to satisfy WMAP data. For larger values of $q$ we can still get
$\omega^{(0)} = -1$ even with large $x_{c}$ ({\it i.e.} the
duration of changeover is small). From Fig. \ref{fig:figw0} we
show the minimum value of $q$ ($q_{\rm{min}}$) for each $x_{c}$ to
satisfy WMAP data ($\omega^{(0)} < -0.78$). For a smaller value of
$x_{c}$ we allow a smaller value of $q_{\rm{min}}$ to satisfy WMAP
due to the fact that $\omega$ changes early and it takes more time
for $\omega$ to reach today when $x_{c}$ is small. This is shown
in Fig. \ref{fig:figw0}(a). We also show the restriction of $q$
for a specific value of $x_{c}$ in Fig. \ref{fig:figw0}(b). The
advantages of this parametrization are its simplicity and the
possibility of mimicking various models. However this
parametrization cannot account for an oscillating $\omega$, which
can be obtained from a $\cosh (\phi)$-potential and the coupled
cases with general potentials \cite{Sahni, LOP}. Thus this
parametrization is good only for monotonically varying $\omega$.

\section{CMB peak spacing }
\setcounter{equation}{0}

The location of the acoustic peaks and the spacing between the
peaks can be estimated if an adiabatic initial condition and a
flat universe are assumed \cite{QCMB, HuDoran}. The acoustic
scale, $l_{A}$ is given by the simple formula \be l_{A} \equiv \pi
\frac{d_{A}}{s} = \pi \frac{\tau_{0} - \tau_{ls}}{s} = \pi
\frac{\tau_{0} - \tau_{ls}}{\bar{c}_{s} \tau_{ls}} \label{deltal}
\ee where $d_{A}$ is the angular size distance to decoupling,
$\tau_{0}$ and $\tau_{ls}$ are the conformal time today and at
last scattering. The sound horizon at decoupling is given by $s =
\bar{c}_{s} \tau_{ls}$ and the average sound speed before last
scattering is, \be \bar{c}_{s} = \frac{\int_{0}^{\tau_{ls}} c_{s}
d \tau}{\tau_{ls}} ~~{\rm where} ~~ c_{s}^{-2} = 3 + \frac{9}{4}
\frac{\rho_{b}(t)}{\rho_{r}(t)} 
\label{cs} \ee where $\rho_{b}$ and $\rho_{r}$ are the baryon and
photon energy densities, respectively. The location of the $m$-th
peak and trough is slightly shifted by driving effects and this
can be written as \cite{Doran} ; \be l_{m} \equiv l_{A} (m -
\varphi_{m}) \equiv l_{A} (m - \bar{\varphi} - \delta \varphi_{m})
\label{lm} \ee where $\bar{\varphi} \equiv \varphi_{1}$ is the
overall peak shift and $\delta \varphi_{m}$ is the shift of the
$m$-th peak relative to the first. We attach the fitting formulae
in the appendix for the sake of completeness. To see the effect of
quintessence, we start from the Friedmann equation. \be H^2 =
\frac{1}{3 \bar{M}^2} (\rho_{\phi} + \rho_{r} + \rho_{m}) \equiv
\frac{1}{3 \bar{M}^2} \rho_{cr} \label{H} \ee where $H$ is the
Hubble expansion rate, $\bar{M}^2 = M_{p}^2/8 \pi$ is a reduced
Planck mass, and $\rho_{i}$ are the energy densities of each
component. From the continuity equation of energy density we find,
\be d \ln \rho_{\phi} = - 3 (1 + \omega) dx. \label{drho} \ee This
equation can be integrated using the parametrization of $\omega$
in Eq. (\ref{omega}) : \be \rho_{\phi}(x) = \rho_{\phi}^{(0)} \exp
(-4x) \Biggl( \frac{ \exp (qx) + \exp (qx_{c})}{1 + \exp (qx_{c})}
\Biggr)^{4/q} = \rho_{\phi}^{(0)} a^{-4} \Biggl( \frac{ a^{q} +
a_{c}^{q}}{1 + a_{c}^q} \Biggr)^{4/q} \label{rho} \ee where
$\rho_{\phi}^{(0)}$ is the dark energy density today. We also find
the dark energy density contrast ($\Omega_{\phi}$) from the above
equation (\ref{rho}). \be \Omega_{\phi}(a) =
\frac{\rho_{\phi}}{\rho_{cr}} = \Biggl[ 1 +
\frac{\Omega_{m}^{(0)}}{\Omega_{\phi}^{(0)}} (a + a_{eq}) \Biggl(
\frac{ a^{q} + a_{c}^{q}}{1 + a_{c}^q} \Biggr)^{-4/q} \Biggr]^{-1}
\label{Omegaphi} \ee  where $\Omega_{i}^{(0)} =
\rho_{i}^{(0)}/\rho_{cr}^{(0)}$ is the present energy density
contrast of each component. In the second equality we use the
relation $\Omega_{r}^{(0)} = \Omega_{m}^{(0)} a_{eq}$ where
$a_{eq}$ is the scale factor when the radiation and matter
densities are equal.
\begin{table}[htb]
\begin{center}
\caption{Cosmological parameters used in the analysis. We use WMAP
data \cite{Spergel}. \label{tab:11}} \vskip .3cm
\begin{tabular}{|c|c|c|c|c|c|}
\hline $\Omega_{\phi}^{(0)}$& $\Omega_{m}^{(0)}$ &
$\Omega_{b}^{(0)} h^2$ & $z_{ls}$ & $z_{eq}$ & $n_{s}$ \\
\hline $0.73$& $0.27$ & $0.0224$ & $1089$ & $3233$& $0.93$ \\
\hline
\end{tabular}
\end{center}
\end{table}
\begin{center}
\begin{figure}
\vspace{1cm} \centerline{ \psfig{file=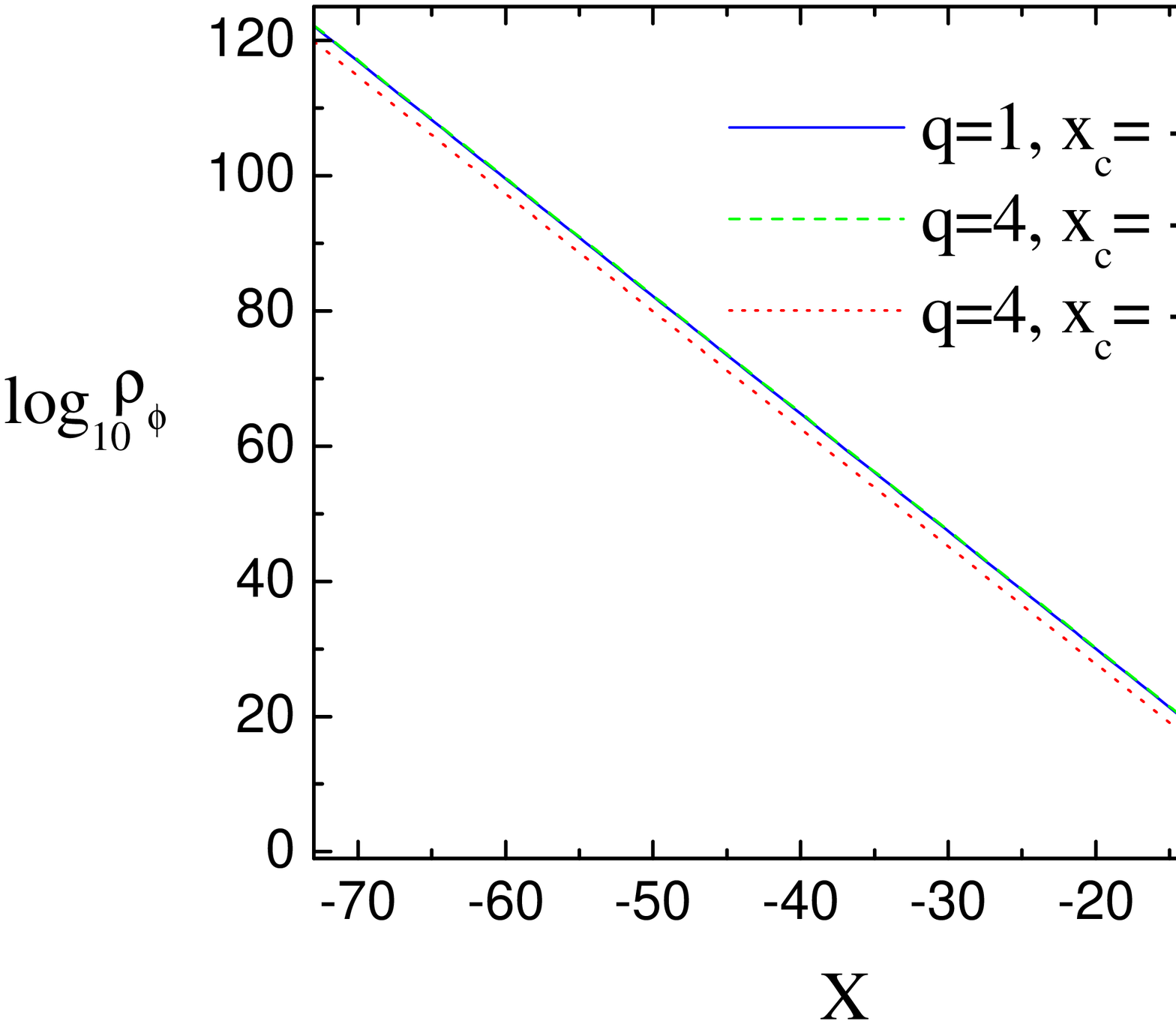, width=8cm}
\psfig{file=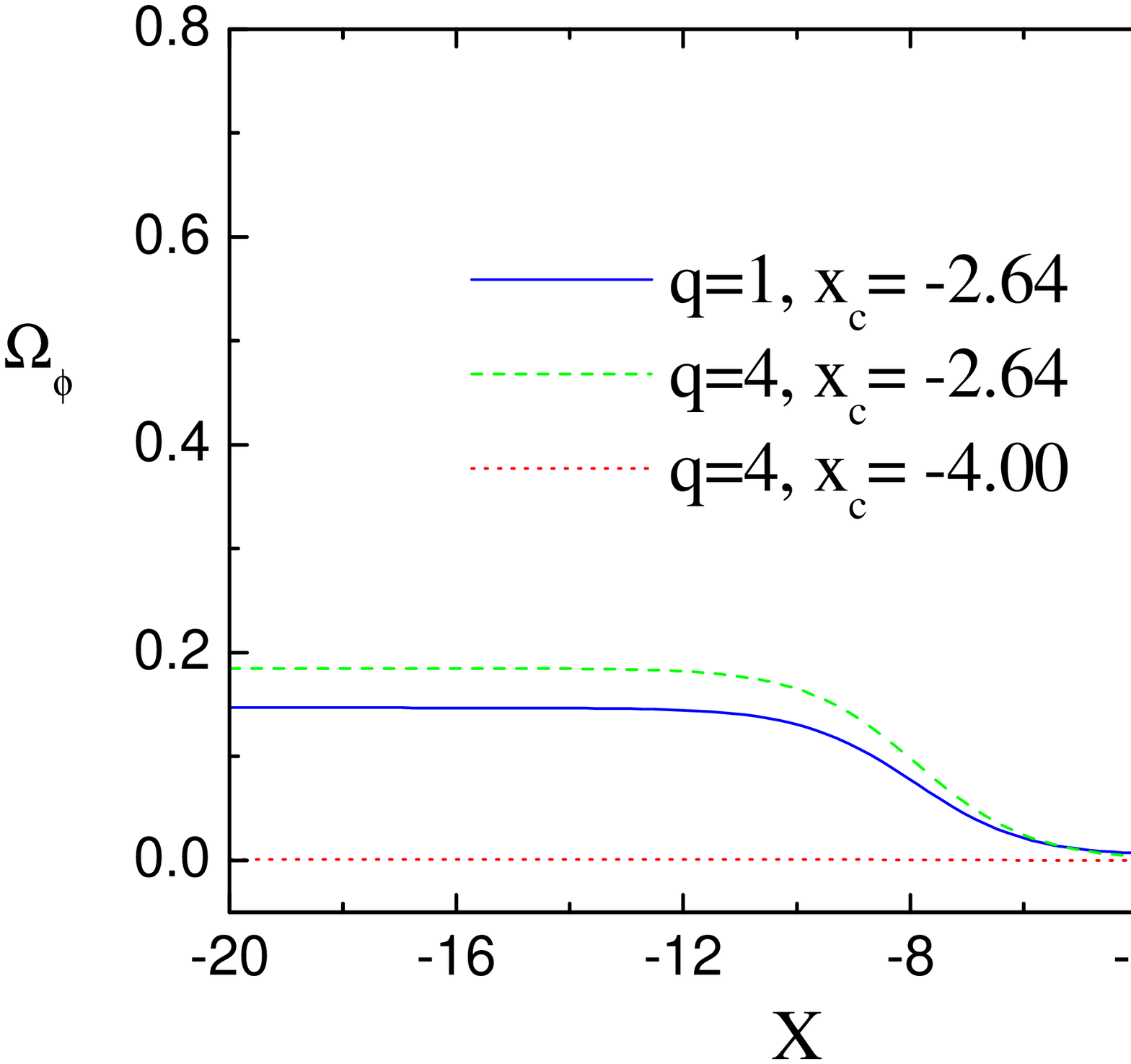, width=8cm} } \vspace{-2cm} \caption{ (a) The
cosmological evolution of $\rho_{\phi}$ for specific $q$ and
$x_{c}$ values. For other values of these parameters, we can get
similar behaviors which can account for $120$ orders of magnitude
change in $\rho_{\phi}$. (b) The evolution of $\Omega_{\phi}$ when
we fix $\Omega_{\phi}^{(0)} = 0.73$. When $x_{c} = -2.64$, a
smaller value of $q = 1$ (solid line) shows a slower change in
$\Omega_{\phi}$ at late universe compared to $q = 4$ (dashed
line). For a smaller value of $x_{c} = -4.00$ (dotted line),
$\Omega_{\phi}$ is negligible during early universe for any value
of $q$.} \label{fig:fig2}
\end{figure}
\end{center}
From Eq. (\ref{rho}), we can see that the dark energy density,
$\rho_{\phi}$ evolves as the energy density of radiation
($\rho_{r} \propto a^{-4}$) during $a \ll a_{c}$. $\rho_{\phi}$
starts to departure from this behavior as $a$ approaches $a_{c}$.
This explains the tracking behavior of dark energy during the
radiation dominated era. The cosmological evolution of the dark
energy density is shown in Fig. \ref{fig:fig2}. $\rho_{\phi}$ is
changed by $120$ orders of magnitude through the entire history of
the universe. Also from Eq. (\ref{rho}) we know that the initial
value of dark energy, $\rho_{\phi}^{(0)} (a_{c}/a_{i})^4$ is
independent of $q$, where $a_{i}$ is the scale factor at the
beginning of the universe. If we use the fact that
$\rho_{\phi}^{i}/ \rho_{\phi}^{(0)} \sim 10^{120}$ and $a_{i} \sim
10^{-32}$, then we find that $a_{c}$ should be approximately
$10^{-2}$ (equally $x_{c} \sim -4$). We can make further
restriction on $a_{c}$ from the following consideration. To be
compatible with observational data, the energy density of
quintessence must be subdominant during Big Bang Nucleosynthesis
(BBN) \cite{FJ}, $\Omega_{\phi}^{(BBN)}(x \sim -23) \leq 0.2$ at
$T \sim 1$ MeV. If we consider the fact that $1 \gg a_{c} \gg
a_{eq} \gg a_{BBN}$, then the equation of dark energy density
contrast (\ref{Omegaphi}) is approximately :

\be \Omega_{\phi}(a_{BBN}) \simeq \Biggl[ 1 +
\frac{\Omega_{m}^{(0)}}{\Omega_{\phi}^{(0)}}
\frac{a_{eq}}{a_{c}^4} \Biggr]^{-1} \leq 0.2 \Longrightarrow a_{c}
\simeq \Biggl( \frac{1 - \Omega_{\phi}^{(0)}}{\Omega_{\phi}^{(0)}}
\frac{\Omega_{\phi}^{(BBN)}}{1 - \Omega_{\phi}^{(BBN)}} a_{eq}
\Biggr)^{1/4} \leq 0.073 \label{Omegaphi1} \ee This corresponds to
$x_{c} \leq -2.60$ independent of $q$. As previously mentioned,
the slope of $\rho_{\phi}$ during the radiation dominated era has
nothing to do with $a_{c}$ or $q$. The effect of $q$ appears when
$a$ approaches $a_{c}$. This effect on $\rho_{\phi}$ is hardly
seen as in Fig. \ref{fig:fig2}(a) due to degeneracy. However we
can see this effect from Fig. \ref{fig:fig2}(b). A larger value of
$q$ (dashed line) gives the steeper change in $\rho_{\phi}$ and
$\Omega_{\phi}$ than those of a smaller value of $q$ (solid line)
at late time. As we can see in the equation (\ref{Omegaphi1}), the
$\Omega_{\phi}$ value depends only on $a_{c}$ before $a$ reaches
to $a_{eq}$. Thus for the smaller value of $a_c$, $\Omega_{\phi}$
value will be more close to the value of cosmological constant,
$\Omega_{\Lambda}$. The evolution of dark energy density shows
degeneracy between the different choice of parameters on $\omega$.
This may require additional experimental data to investigate the
evolution of $\omega$ instead of using SNe. Because SNe data is
analyzed by the luminosity distance, which mainly depends on
$\rho_{\phi}$. In the above graphs we fix $\Omega_{\phi}^{(0)} =
0.73$. Now we rewrite the equation (\ref{H}) by using conformal
time, $\tau$ : \ba \Biggl(\frac{d a}{d \tau}\Biggr)^2 &=& H_{0}^2
\Biggl\{ \Omega_{r}^{(0)} + \Omega_{m}^{(0)} a +
\Omega_{\phi}^{(0)} \Bigl(\frac{a^{q} + a_{c}^q}{1 + a_{c}^q}
\Bigr)^{4/q} \Biggr\} = H_{0}^2 \Biggl\{ \Omega_{m}^{(0)} (a +
a_{eq}) + \Omega_{\phi}^{(0)} \Bigl(\frac{a^{q} + a_{c}^q}{1 +
a_{c}^q} \Bigr)^{4/q} \Biggr\} \nonumber \\ &=& H_{0}^2 \Biggl\{
(1 -\Omega_{r}^{(0)} - \Omega_{\phi}^{(0)} ) ( a + a_{eq}) +
\Omega_{\phi}^{(0)} \Bigl(\frac{a^{q} + a_{c}^q}{1 + a_{c}^q}
\Bigr)^{4/q} \Biggr\} \label{da1} \ea where $H_{0}$ is the present
value of Hubble parameter. If we use WMAP data $\Omega_{r}^{(0)}
\simeq 4.9 \times 10^{-5}$, $\Omega_{m}^{(0)} \simeq 0.27$,
$\Omega_{\phi}^{(0)} \simeq 0.73$, and $a_{eq} = 1/(1+z_{eq}) =
1/3234$, then we have the following approximation of the above
equation (\ref{da1}). \be \Biggl(\frac{d a}{d \tau}\Biggr)^2
\simeq H_{0}^2 \Biggl\{ (1 - \Omega_{\phi}^{(0)}) (a + a_{eq}) +
\Omega_{\phi}^{(0)} \Bigl(\frac{a^{q} + a_{c}^q}{1 + a_{c}^q}
\Bigr)^{4/q} \Biggr\} \neq H_{0}^2 \Biggl\{ (1 -
\Omega_{\phi}^{(0)}) a + \Omega_{\phi}^{(0)} \Bigl(\frac{a^{q} +
a_{c}^q}{1 + a_{c}^q} \Bigr)^{4/q} \Biggr\} \label{da2} \ee The
approximation comes from the fact that $\Omega_{r}^{(0)} \ll
\Omega_{m, \phi}^{(0)}$. We indicate the inequality between the
first and the second expressions, which is frequently used in the
literature. The numerical difference between two expressions is
not negligible for the specific values of $a_{c}$ and $q$. So we
will use the first (accurate) expression in the following
calculations. From this equation we can find the numerical value
of $\tau$ when $a_{c}$ and $q$ are fixed. \be d \tau = H_{0}^{-1}
\Biggl( (1 - \Omega_{\phi}^{(0)}) (a + a_{eq}) +
\Omega_{\phi}^{(0)} \Bigl(\frac{a^{q} + a_{c}^q}{1 + a_{c}^q}
\Bigr)^{4/q} \Biggr)^{-1/2} da \label{dtau} \ee

\begin{center}
\begin{figure}
\vspace{1cm} \epsfxsize=5.5cm \centerline{\psfig{file=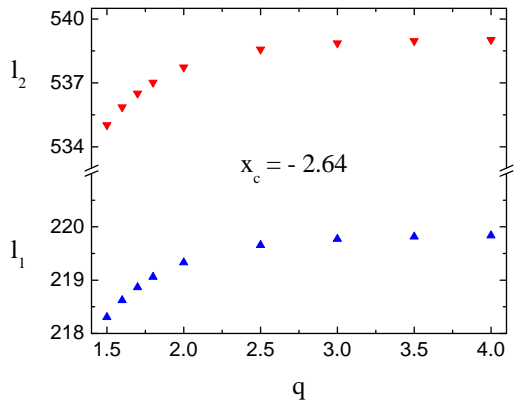,
height=6cm}} \vspace{-2cm} \caption{The locations of the first and
the second CMB peaks. The uptriangles indicate the first peaks,
$l_{1}$ and the downtriangles show the second peaks, $l_{2}$ as a
function of $q$. Both $l_{1}$ and $l_{2}$ converge as $q$
increases.} \label{fig:fig5}
\end{figure}
\end{center}

\begin{table}[htb]
\begin{center}
\caption{The spacing and the location of the CMB peaks for several
values of $q$ when $x_{c} = -2.64$. Here we use the cosmological
parameters in Table \ref{tab:11}. \label{tab:12}} \vskip .3cm
\begin{tabular}{|c|c|c|c|c|c|}
\hline $x_{c}$& $q$ & $l_{A}$ & $l_{1}$ &$l_{2}$&
$\Omega_{\phi}^{ls}(10^{-2})$ \\
\hline $-2.64$& $1.5$ & $300.8$ & $218.3$ & $535.0$& $5.17$ \\
&$2$& $302.4$& $219.3$& $537.7$& $5.36$ \\
&$3$& $303.1$& $219.8$& $538.9$& $5.40$\\
&$4$& $303.1$& $219.8$& $539.0$& $5.41$ \\
\hline
\end{tabular}
\end{center}
\end{table}
By using the above equations we can find the acoustic scale
($l_{A}$), the locations of the first two peaks ($l_{1}, l_{2}$),
and $\Omega_{\phi}$ at last scattering ($\Omega_{\phi}^{ls}$).
They are given in Table \ref{tab:12} and Fig. \ref{fig:fig5}. For
a value of $x_{c}$, the location of CMB peaks depends on
$q$-values. However CMB peak values converge to some values as $q$
increases. From WMAP measurements the acoustic scale and the
locations of the first two peaks are \cite{Spergel} : \be l_{A} =
301 \pm 1, \hspace{0.2in} l_{1} = 220.1 \pm 0.8, \hspace{0.2in}
l_{2} = 546 \pm 10. \label{l12} \ee
\begin{center}
\begin{figure}
\vspace{1cm} \epsfxsize=5.5cm \centerline{\psfig{file=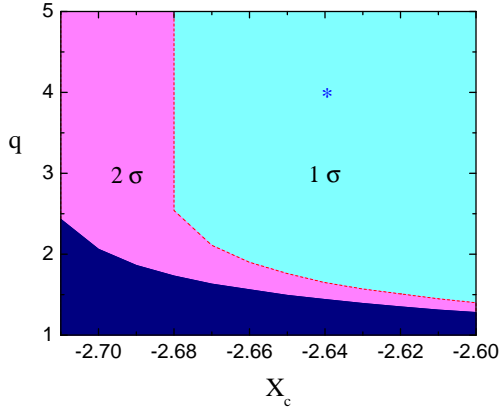,
height=6cm}} \vspace{-2cm} \caption{The $68$\,\% (light shaded)
and 95\,\% (medium shaded) confidence allowed regions for $x_{c}$
and $q$ using WMAP data. Asterisk indicates the best fit values
($x_c = -2.64$, $q=4.01$). The dark shaded regions are out of $2
\sigma$ confidence level. The constraint from the dark energy
density at BBN determines the upper limit on $x_{c} \, (\leq
-2.60)$. The lower limit of $x_{c} \, (> -2.72)$ is obtained from
$\chi^2$-analysis. There is no upper limit on $q$ values from
WMAP. However we can obtain the upper limit on $q$ from the
analysis of the time varying $\alpha$ in the following section.}
\label{fig:chi}
\end{figure}
\end{center}
If we assume that the errors are Gaussian and uncorrelated, then
we can find a $\chi^2$-statistic \be \chi^2 = \sum_{i}
\frac{\Bigl(l_{i, obs} - l_{i, theory}\Bigr)^2}{\sigma_{i}^2}
\label{chi2} \ee where $l_{i,obs}$ is the observational data of
each peak, $l_{i,theory}$ is the theoretical value of $l_{i}$, and
$\sigma_{i}$ is the statistical uncertainty for each peak. The
likelihoods for the parameters ($q, x_{c}$) of the parametrization
given in (\ref{omega}) are calculated from $\chi^2$ so that the
$68$\,\% and $95$\,\% confidence regions are determined by $\Delta
\chi^2 \equiv \chi^2 - \chi_{0}^2 = 2.30$ and $6.17$ respectively.
$\chi_{0}^2$ is $\chi^2$ for the best fit model found. The best
fit values to the WMAP data is obtained from $x_{c} = -2.64$ and
$q = 4.01$. The likelihood contours for WMAP are shown in Fig.
\ref{fig:chi}. The $68$\,\% (light shaded) and 95\,\% (medium
shaded) confidence allowed regions for $x_{c}$ and $q$ are shown
using WMAP data. There is no upper limit on $q$ values for several
$x_{c}$ values. From this we find at $95$\,\% CL : \be
\omega^{(0)} \leq -0.971_{-0.027}^{+0.017}~, \hspace{0.2in}
\Biggl( \frac{d \omega}{d z} \Biggr)_{z=0} \leq
0.041_{-0.037}^{+0.016} \label{wdw0} \ee $\omega$ is almost a
constant at present, which is consistent with recent observation
\cite{Riess}.


\section{Time variation of the fine structure constant }
\setcounter{equation}{0}
In this section we consider the $\phi$-dependence of the gauge
couplings in Standard Model (SM), $B_{F}(\phi)$ \cite{LOP}. The
interaction of a light scalar field $\phi$ with electromagnetic
field is represented in the Lagrangian as ${\cal L}_{EM} =
-B_{F}(\phi) F_{\mu\nu} F^{\mu\nu} /4$, where $F^{\mu\nu}$ is the
electromagnetic field-strength tensor. The main assumption in this
work is that all functions $B_{F}(\phi)$ and $V(\phi)$ admit a
common extremum, which is a generalization of Damour-Nordtvedt and
Damour-Polyakov constructions \cite{DNP}. To this end, we proposed
an ansatz that would relate $B_{F}(\phi)$ and $V(\phi)$. \be
B_{F}(\phi) = \Biggl(\frac{b_{F} + V(\phi)/V_{0}}{1 + b_{F}}
\Biggr)^{n_{F}} , ~~~ {\rm where}~~ b_{F} + 1 > 0 \label{B1} \ee
where $n_{F}$ and $b_{F}$ are two dimensionless parameters
allowing us to cover a wide range of possibilities. We will assume
that dark energy is a quintessence field and its potential is
parameterized by $\omega$. So instead of introducing a potential
of the scalar field, we will use $\omega$. We can rewrite
(\ref{B1}) by using the parametrization of $\omega$ (\ref{omega}).
We also use the relation between the potential $V(\phi)$ and the
energy density $\rho_{\phi}$ of the scalar field, \be V(\phi) =
\frac{(1 - \omega)}{2} \rho_{\phi} \Rightarrow V(x) =
\rho_{\phi}^{(0)} \exp (-4x) A_{1}(x) A_{2}(x) \label{V} \ee where
\ba A_{1}(x) &=& \Biggl( \frac{ \exp (qx) + \exp (qx_{c})}{ 1 +
\exp (qx_{c})} \Biggr)^{4/q} ~ ,
\label{A1} \\
A_{2}(x) &=& \frac{ 3 \exp (qx) + \exp (qx_{c})}{3 \Bigl( \exp
(qx) + \exp (qx_{c}) \Bigr)} \label{A2} \ea

\begin{center}
\begin{figure}
\vspace{1cm} \centerline{ \psfig{file=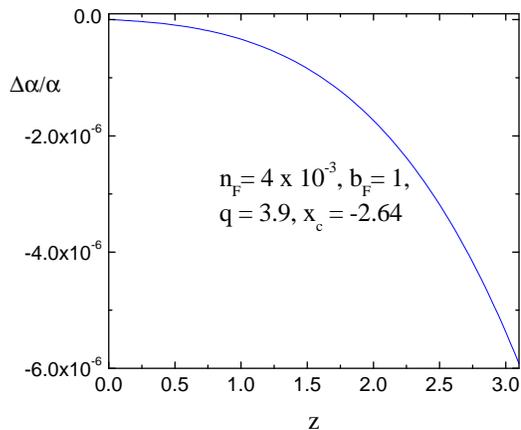,
width=8cm} } \vspace{-2cm} \caption{ A typical evolution $\Delta
\alpha / \alpha$ for specific $n_F$, $b_F$, $q$ and $x_{c}$
values. We can get similar behaviors with other values of
parameters. $n_F$ and $b_F$ values are obtained from the result of
Murphy {\it et al}. Several values of $\Delta \alpha / \alpha$ for
various choices of the parameters at different $z$ are indicated
in Table \ref{tab:21}.} \label{fig:fig3}
\end{figure}
\end{center}
With these equations (\ref{omega}), (\ref{rho}), and (\ref{V}), we
can rewrite the gauge coupling in Eq (\ref{B1}) as follow, \be
B_{F}(x) = \Biggl(\frac{b_{F} + \exp (-4x) A_{1}(x) A_{2}(x) /
A_{2}(0)}{1 + b_{F}} \Biggr)^{n_{F}} \label{B2} \ee where we
assume that $V_{0} = V(x=0)$ to satisfy $B_{F}(0) = 1$. We can
find the time variation of the fine structure constant by using
the following relation between $B_{F}$ and $\alpha$, \be
\frac{\Delta \alpha}{\alpha} \equiv \frac{\alpha(x) -
\alpha(0)}{\alpha(0)} = \frac{1}{B_{F}(x)} -1. \label{deltaalpha1}
\ee From equations (\ref{B2}) and (\ref{deltaalpha1}), we find \be
\frac{\Delta \alpha}{\alpha} = \Biggl( \frac{ 1 + b_{F} }{b_{F} +
\exp (-4x) A_{1}(x) A_{2}(x) / A_{2}(0)} \Biggr)^{n_{F}} - 1
\label{deltaalpha2} \ee

\begin{table}[htb]
\begin{center}
\caption{The values of $\Delta \alpha / \alpha$ at the different
$z$ for the several values of $q$ when $x_{c} = -2.64$. Here we
use the QSOs result to fix other values \cite{Webb}. $n_{F}$ has
been scaled by a factor of $10^{3}$, $\Delta \alpha / \alpha$
values have been scaled by a factor of $10^{6}$ except the value
at decoupling $(\Delta \alpha / \alpha)_{ls}$, which is given in
the last column without any scaling. \label{tab:21}} \vskip .3cm
\begin{tabular}{|c|c|c|c|c|c|c|c|c|}
\hline $x_{c}$& $q$ & $n_{F}$ & $b_{F}$ &
$(\frac{\Delta\alpha}{\alpha})_{3}$ &
$(\frac{\Delta\alpha}{\alpha})_{1.5}$ &
$(\frac{\overline{\Delta\alpha}}{\alpha})_{0.45}$ &
$(\frac{\Delta\alpha}{\alpha})_{0.14}$ &
$(\frac{\Delta\alpha}{\alpha})_{1089}$ \\
\hline $-2.64$& $3$ & $4.25$ & $11$ &
$-5.4$& $-1.25$ & $-0.017$ & $-0.041$ & $-0.057$ \\
& & $0.71$& $1$ & $-5.4$& $-1.26$ & $-0.017$& $-0.042$ & $-0.011$ \\
\hline
&$3.5$& $3.68$& $3$& $-5.4$& $-1.01$ &$-0.023$ & $-0.025$ & $-0.053$  \\
& & $1.84$& $1$& $-5.4$& $-1.01$ & $-0.023$ & $-0.025$& $-0.028$  \\
\hline
&$3.9$& $4.01$& $1$& $-5.4$& $-0.84$ &$-0.016$ & $-0.016$ & $-0.060$  \\
\hline
\end{tabular}
\end{center}
\end{table}
\begin{center}
\begin{figure}
\vspace{1cm} \epsfxsize=5.5cm
\centerline{\psfig{file=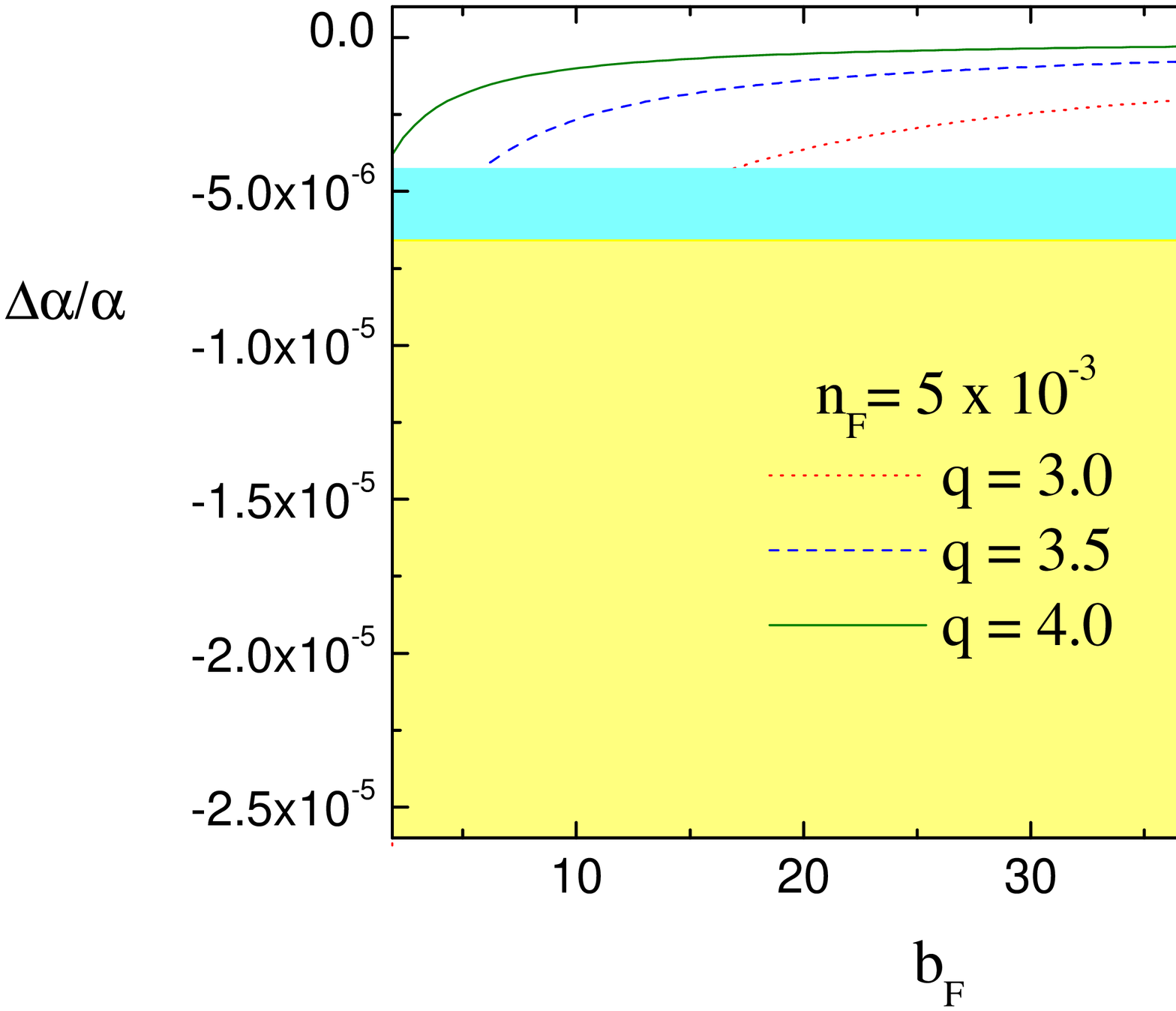,
height=6cm}\psfig{file=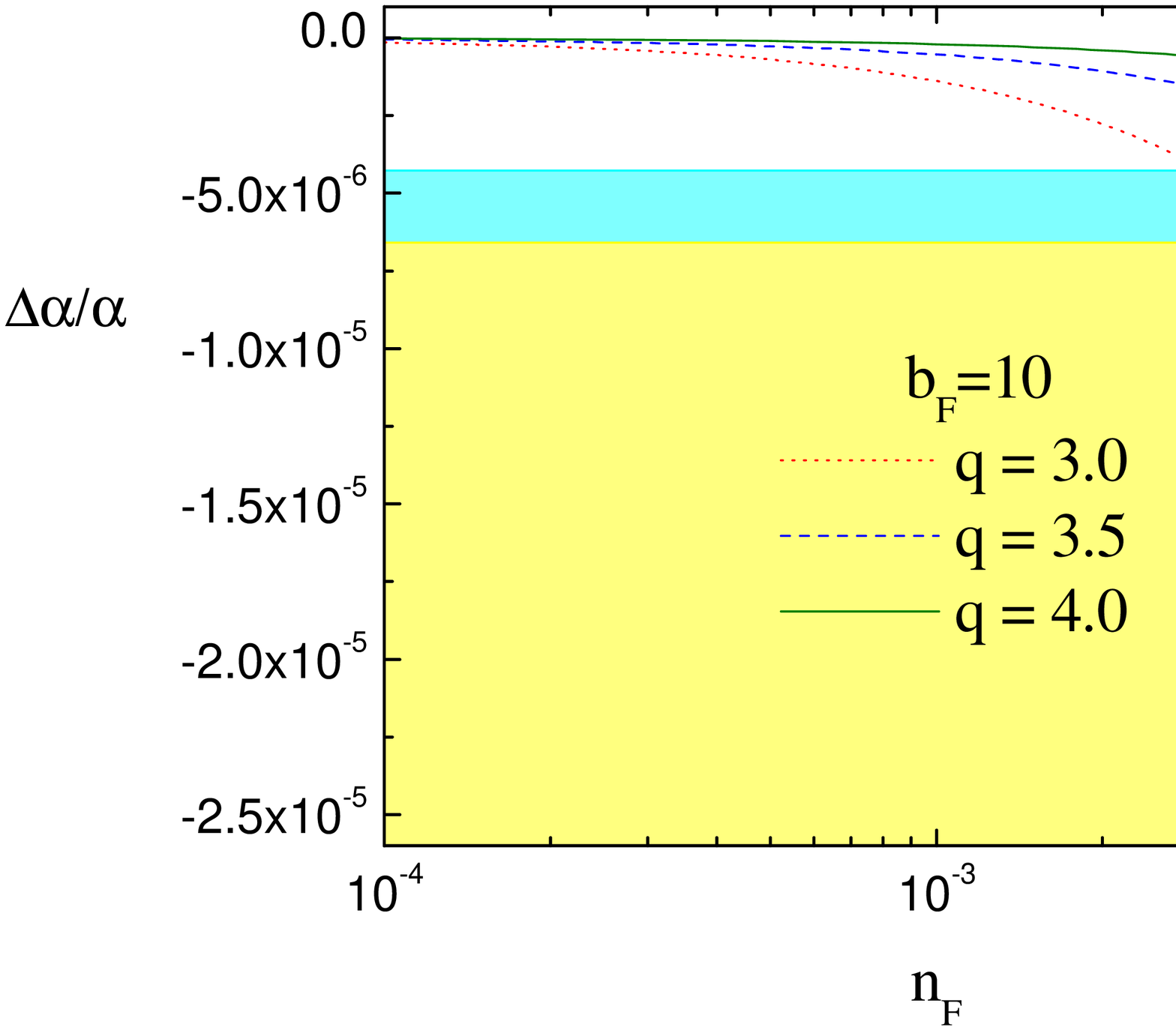, height=6cm}}
\epsfxsize=5.5cm \centerline{\psfig{file=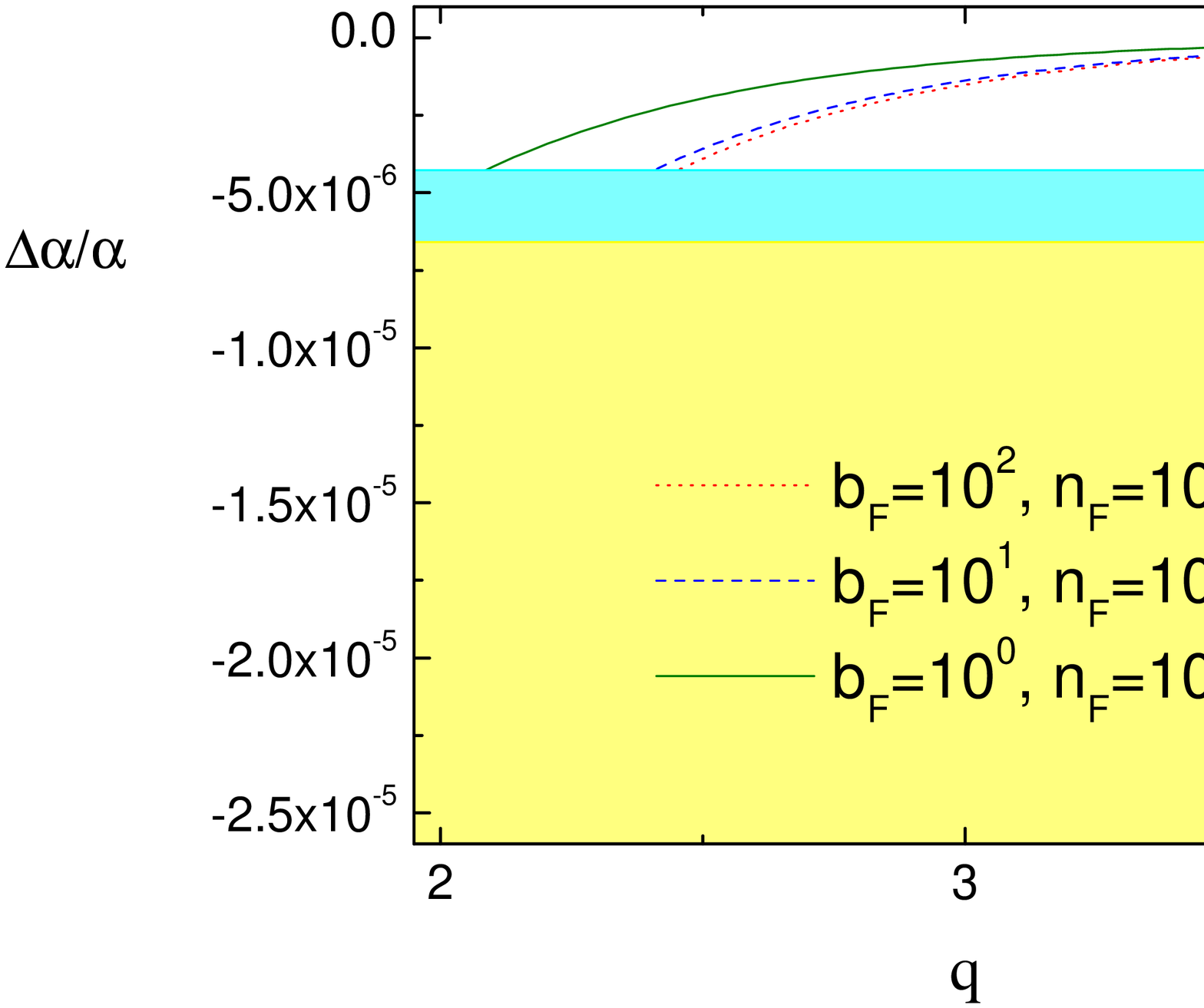,
height=6cm}\psfig{file=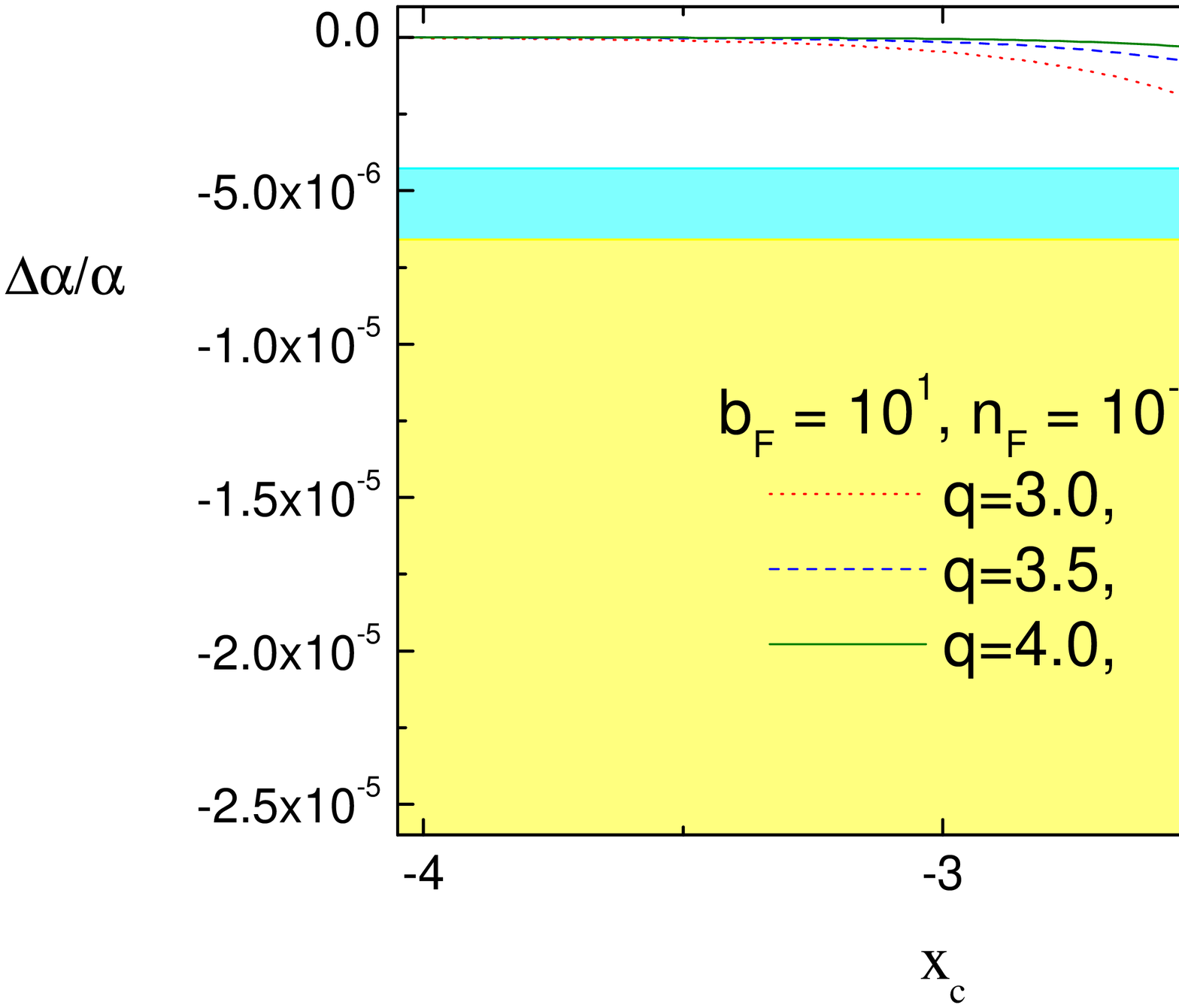, height=6cm}} \vspace{-2cm}
\caption{ The cosmological evolution of $\alpha$ for the various
parameters. Murphy {\it et al} data (medium shaded) is used in all
of the figures, $(\Delta \alpha/ \alpha)_{z=3} = (-0.543 \pm
0.116) \times 10^{-6}$. We use $x_{c} = -2.64$ for all of the
graphs except (d). (a) The evolution of $\Delta\alpha/\alpha$ with
respect to $b_{F} \geq 1$ for $n_{F} = 5 \times 10^{-3}$. When $q
=3$ (dotted line) $\alpha$ changes slower than when $q = 4$ (solid
line). Thus we have more acceptable values of $b$ for the smaller
$q$ ($10.6 \leq b_{F} \leq 16.9$ when $q = 3$). (b) The effect of
changing $n_{F}$ on $\Delta \alpha / \alpha$ while we fix $b_{F} =
10$. (c) and (d) the effect of varying $q$ and $x_{c}$ on the
variation of $\alpha$, respectively.} \label{fig:fig4}
\end{figure}
\end{center}
In Table \ref{tab:21}, we display $\Delta \alpha / \alpha$ values
at different redshifts for the various parameters. Especially we
show specific results for $x_{c} = -2.64$, because this gives the
best-fit value for the separation of CMB peaks as explained in the
previous section. $\Delta \alpha / \alpha$ at the cosmic microwave
background epoch, which is constrained as $(\Delta \alpha /
\alpha)_{ls} \geq -0.06$ \cite{cmb} are shown in the last column.
There are four parameters ($n_{F}, b_{F}, x_{c}$, and $q$) to be
fixed. In Fig. \ref{fig:fig3} we show a typical evolution of
$\Delta \alpha / \alpha$ for a specific set of four parameters
({\it i.e.} $n_{F} = 4 \times 10^{-3}$, $b_{F} =1$, $q=3.9$, and
$x_{c} =-2.64$). We can change four parameters to explain the
effect on the evolution of $\alpha$. However it is impossible to
fix all four parameters at once. Thus we show the effect of the
changing each parameter with fixing all other parameters at each
time. We see that smaller values of $q$ lead to more room for the
parameter space regions of $n_{F}$ and $b_{F}$, because smaller
values of $q$ lead to slower change in $\alpha$. For given
$q$-values, there may not be any solution to fit experimental data
as shown in the first panel of Fig. \ref{fig:fig4}. However as we
increase the $b_{F}$-values we can satisfy the experimental data.
We have larger $b_{F}$ range for smaller values of $q$. Similar
consideration can be made for varying $n_{F}$. For example, even
though the dotted lines (q = 3) in the second panel does not fit
experimental data when $n_{F} \geq 3 \times 10^{-3}$, as we
decrease $n_{F}$ we can recover the suitable $\Delta \alpha /
\alpha$ for given $q$ and $x_{c}$. The medium shaded regions of
each panel are allowed from Murphy {\it et al}, ($\Delta \alpha /
\alpha)_{z=3} = (-0.543 \pm 0.116) \times 10^{-5}$ \cite{Webb}.
The effect of changing $q$ and $x_{c}$ are shown in $(c)$ and
$(d)$ of Fig. \ref{fig:fig4}. As we expect there is lots of room
for the choice of $q$ and $x_{c}$ because we have extra degrees of
freedom ($n_{F}$ and $b_{F}$) to be fixed in addition to the
original parameters of $\omega$. However combined with the
analysis of the separation of CMB peaks, the parameters on gauge
coupling are much narrowed as in Table \ref{tab:21}. We can make
the allowed regions for $n_{F}$ and $b_{F}$ from the narrowed
parameter spaces of $a_c$ and $q$. This is indicated in Fig.
\ref{fig:fig6}. The sloped lines are obtained from the
normalization of Murphy {\it et al} data for the given $q$ values
when $x_{c}$ is fixed as $-2.64$. The vertical lines for each $q$
come from both the CMB bound on $\alpha$ and the assumption of $b
\geq 1$. There is only one allowed value $b_{F} = 1$ for $q = 3.9$
( solid line ). This gives the upper limit on $q \leq -3.9$, which
we can use for the consideration of CMB peaks. For $q =3$ (dotted
line), there are more parameter spaces to satisfy the constraints.
In this case $b_{F}$ can be increased up to $11$. Obviously this
gives much narrower parameter spaces for each parameter compared
to Fig. \ref{fig:fig4}. Now we can restrict ($n_{F},b_{F}$) $\sim$
($10^{-3}, 1$) as in \cite{LOP} and $1.5 \leq q \leq 3.9$ for
$x_{c} = -2.64$.
\begin{center}
\begin{figure}
\vspace{1cm} \epsfxsize=5.5cm \centerline{\psfig{file=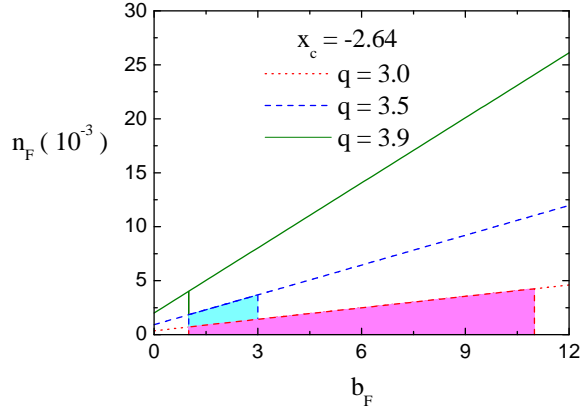,
height=6cm}} \vspace{-2cm} \caption{The constraints on $n_{F}$ and
$b_{F}$ from both the result of Murphy {\it et al} \cite{Webb} and
that of CMB \cite{cmb}. $n_{F}$ values in the graph are scaled by
$10^{3}$. The sloped lines are obtained from QSOs result. The
vertical line for each case is obtained from CMB. The shaded
regions are allowed ones from both constraints. There is only one
$b_F$-value for $q = 3.9$ (solid line) when we assume that $b \geq
1$. So we can get the upper limit $q < 4$.} \label{fig:fig6}
\end{figure}
\end{center}

\section{Conclusions}
\setcounter{equation}{0}

 We have introduced a simple parametrization of $\omega$ based on
both theory and experiment. A tracking solution is proposed to
alleviate the fine tuning problem and the present value of
$\omega$ is observed to be close to $-1$. We have analyzed the
separation of CMB peaks and the time variation of the fine
structure constant with this parametrization. We have obtained the
constraints of the parameters of $\omega$ from these analyses
which use both low-redshifts and high-redshifts data. We have been
able to use the analytically integrated value of the dark energy
density contrast, $\Omega_{\phi}$ from this parametrization.

 From WMAP data, we have obtained a very specific value of $x_{c}
\sim -2.64$ (equally $z_{c} \sim 13$) for the best fit value. When
$x_c = -2.64$ we have obtained the very small $(dw/dz)_{z=0}$,
which is consistent with the recent observations. $(dw/dz)$
depends on $x_c$, which is related to the epoch when the time
evolution of $\omega$ begins. A smaller value of $x_{c}$ causes an
earlier evolution of $\omega$ compared to a larger value of
$x_{c}$. With a smaller $x_c$, $\rho_{\phi}$ evolves to its
present value $\rho_{\phi}^{(0)}$ at an early epoch and it stays
with this value because $\omega \sim -1$. In this case
$\Omega_{\phi}$ is negligible until it reaches its present value
and that is similar to when we have the cosmological constant as
dark energy. However for some values of $x_{c}$ the value of
$\Omega_{\phi}$ can be significant at early times. This behavior
is quite different from that of a cosmological constant. Thus the
difference between the cosmological constant and quintessence is
obvious for some values of $x_{c}$. Due to CMB consideration we
have focused on $x_{c} = -2.64$ and this has shown the big
discrepancy between quintessence and the cosmological constant. We
have also found that to be consistent with BBN ({\it i.e.}
$\Omega_{\phi}^{(BBN)} \leq 0.18$) we need $q \geq 1.50$ when
$x_{c} = -2.64$. Thus this may give one clue for the nature of
dark energy. We have made the restriction to the value of $q$,
$q_{\rm{min}}$ from the constraint on $x_{c}$. As $x_{c}$
increases, we will get bigger $q_{\rm{min}}$ as in Fig.
\ref{fig:figw0}. We have also found the upper limit on $x_{c}$
when we consider the limit of $\Omega_{\phi}^{(BBN)}$, $x_{c} \leq
-2.60$. We have reached to the lower limit $q \geq 1.50$ when we
used $\Omega_{\phi}^{ls} \sim 0.01$ from WMAP. We have had the
similar lower limit on $q$ from the $\chi^2$-statistic. All of
this consideration gives the unique lower limit of $q \geq 1.50$
when we choose $x_c = -2.64$ However there was no upper limit of
$q$ from the CMB analysis alone.

 We have introduced two more parameters ($n_{F}, b_{F}$) from the
generic form of gauge coupling to analyze the evolution of $\Delta
\alpha / \alpha$. However we had only three parameters to be fixed
after specifying $x_{c}$. We have found the viable parameter
regions for each parameter by using $\Delta \alpha / \alpha$ data.
From this we have found the well known parameter regions of gauge
coupling $(n_{F}, b_{F}) \sim (10^{-3}, 1)$ \cite{LOP}. We have
also found the upper limit $q \leq 3.90$ with this analysis.

We have found a viable parametrization of $\omega$ with the very
specific parameter regions, $x_c = -2.64$ and $ 1.50 \leq q \leq
3.90$ from the analyses of both the separation of CMB peaks and
the time variation of $\alpha$. We have also found that $ -1.00
\leq \omega^{(0)} \leq -0.971_{-0.027}^{+0.017}$ and $
1.76_{-0.42}^{+0.29} \times 10^{-4} \leq (d \omega / dz)_{z=0}
\leq 0.041_{-0.037}^{+0.016}$ at $95$\,\% CL. Thus we may not see
any change of $\omega$ at present. Even though the present values
of $\omega$ and $d \omega / dz$ are almost identical to the values
of the cosmological constant, $\Omega_{\phi}$ evolves from BBN to
the present with significant values. Thus this model has the
different cosmological behaviors compared to $\Lambda$CDM model.
We may need to check the similar parametrization and/or more data
to investigate the nature of dark energy. We may also need to
modify $\omega$ to have more than one changeover which can show a
non-monotonic behavior.

\section{Acknowledgements}
\setcounter{equation}{0}

This work was supported in part by DOE grant DE-FG02-94ER-40823.
We would like to thank Keith.A. Olive for useful discussion.

\appendix
\section{Appendix} \setcounter{equation}{0}

We have used the analytic approximations for the phase shifts
{Doran}, which we show here. The overall phase shift is given by
\be {\bar\varphi}=(1.466-0.466 n_s) \left[a_1 r_*^{a_2} + 0.291
{\bar\Omega}_{\phi}^{ls} \right] \label{phibar} \ee where the
fitting coefficients are \ba
a_1 &=& 0.286+0.626\omega_b \nonumber\\
a_2 &=& 0.1786-6.308\omega_b+174.9\omega_b^2-1168\omega_b^3
\label{as} \ea with $\omega_b=\Omega_b^{(0)} h^2$, and ${\bar
\Omega}_{\phi}^{ls}$ is given by \be {\bar \Omega}_{\phi}^{ls} =
\tau_{ls}^{-1} \int_{0}^{\tau_{ls}} \Omega_{\phi}(\tau) d \tau
\label{Omegabar} \ee and \be r_*\equiv
\rho_{r}(z_{ls})/\rho_{m}(z_{ls}) \ee is the ratio of radiation to
matter at decoupling. The relative shift of the first acoustic
peak is zero, $\delta\varphi_1=0$, and the relative shifts of the
second peak are given by \be
\delta\varphi_2=c_0-c_1r_*-c_2r_*^{-c_3}+0.05(n_s-1)~
\label{delphi2} \ee where $n_s$ is scalar spectral index and \ba
c_0 &=& -0.1+\left(0.213-0.123{\bar\Omega}_{ls}^\phi \right)
\times\exp \Bigl(-[52-63.6 {\bar\Omega}_{ls}^\phi]
\omega_b \Bigl) \nonumber\\
c_1 &=& 0.015+0.063\exp\left(-3500\omega_b^2\right) \nonumber\\
c_2 &=& 6\times 10^{-6}+0.137(\omega_b-0.07)^2 \nonumber\\
c_3 &=& 0.8+ 2.3 {\bar\Omega}_{ls}^\phi
+\left(70-126{\bar\Omega}_{ls}^\phi\right)\omega_b \label{c} \ea


\end{document}